\documentclass[preprint,showpacs,preprintnumbers,amsmath,amssymb]{revtex4}
\usepackage[dvips]{graphicx}
\usepackage{graphicx}
\usepackage{amsfonts}
\usepackage{bm}
\usepackage{amsmath}
\usepackage{amssymb}
\usepackage{color}
\usepackage[all]{xy}

\def\be{\begin{equation}}
\def\ee{\end{equation}}
\def\bea{\begin{eqnarray}}
\def\eea{\end{eqnarray}}

\begin{document}

\title{Geometrothermodynamics of black hole binary systems}

\author{Hernando Quevedo}
\email{quevedo@nucleares.unam.mx}
\affiliation{Instituto de Ciencias Nucleares, Universidad Nacional Aut\'onoma de M\'exico, AP 70543, Ciudad de M\'exico 04510, Mexico}
\affiliation{Dipartimento di Fisica and ICRA, Universit\`a di Roma "Sapienza", I-00185, Roma, Italy}
\affiliation{Institute of Experimental and Theoretical Physics,
    Al-Farabi Kazakh National University, Almaty 050040, Kazakhstan}

\author{Mar\'ia N. Quevedo}
\email{maria.quevedo@unimilitar.edu.co}
\affiliation{Departamento de matem\'aticas, Facultad de Ciencias
B\'asicas,Universidad Militar de Nueva Granada, Cra 11 No. 101-80,
Bogot\'a D.E.,Colombia}

\author{Alberto S\'anchez}
\email{asanchez@ciidet.edu.mx} \affiliation{Departamento de
posgrado, CIIDET,\\{\it AP752}, Quer\'etaro, QRO 76000, MEXICO}

\date{\today}

\begin{abstract}
We study a stationary and axisymmetric  binary system composed of two identical Kerr black holes, whose physical parameters satisfy the Smarr thermodynamic formula. Then, we use the formalism of geometrothermodynamics to show that the spatial distance between the black holes must be considered
as a thermodynamic variable. We investigate the main thermodynamic properties of the system by using the contact structure of
the phase space, which generates the first law of thermodynamics and the equilibrium conditions. The phase transition structure
of the system is investigated through the curvature singularities of the equilibrium space.
It is shown that the thermodynamic and stability properties and the phase transition structure
of the  binary system strongly depend on the distance between the black holes.


\end{abstract}


\maketitle


\section{Introduction}
\label{intro}

One of the main difficulties for understanding the  dynamical
scenario of two rotating black holes are the effects produced by
multipolar interactions between the sources \cite{Cabrera}.
This is
the reason why these systems  are studied by means of
stationary axisymmetric spacetimes with Kerr type sources.
The first work in this direction was published in 1947 and presents a class of static solutions of the
Einstein--Maxwell equations \cite{Majumdar,Papapetrou}.
Later on,  these solutions were generalized to include
the stationary case \cite{Israel1} (see also Ref. \cite{Perjes}).
However, it was not until
1973 that a metric was found, which describes two identical
Kerr--Newman sources in equilibrium under their mutual
electromagnetic and gravitational forces, with their spins
oppositely oriented along a given axis \cite{Parker}. In 1998,
Varzugin \cite{Varzugin} studied a counter-rotating system of two
identical black holes with opposite spin and showed that the
interaction forces are equal compared with two identical
Schwarzschild black holes.
 In recent years, a 3--parametric
physical model was found that  describes  a two black hole system composed by
identical co-rotating Kerr sources separated by a massless strut \cite{Cabrera}.
The solution is characterized by the physical parameters \cite{Komar}
mass $M$, electric charge $Q$, angular momentum $J$, and the distance between the centers
of the black holes  $R$.

On the other hand, the thermodynamic  aspects of rotating black hole binary systems have been also studied in
\cite{Cabrera,Macias,Miguel}, using the Smarr formula \cite{Smarr} and the entropy relation proposed by Bekenstein \cite{Bekenstein}.
An alternative approach consists in using the geometric properties of the equilibrium space of the system under consideration.
In this connection, there are essentially two methods; the first one, called thermodynamic geometry, consists in introducing into the equilibrium space Hessian metrics, which depend on the choice of thermodynamic potential  \cite{weibook,ruprev},  and the second one, called geometrothermodynamics (GTD) \cite{quev07}, uses Legendre invariant metrics instead, i.e., metrics which do not depend on the choice of thermodynamic potential.

In this work, we analyze a binary system which is composed of two identical Kerr black holes. The sources are located along a symmetry axis, which is also the rotational axis of the black holes, and the entire system is assumed to be stationary. The thermodynamic properties of the system are investigated by using entirely the formalism of GTD. First, we consider the contact structure of the phase space which, when projected on the equilibrium space, leads to the first law of thermodynamics and the equilibrium conditions (equations of state). This approach allows us to investigate the behavior of all extensive and intensive variables from  the knowledge of the fundamental equation, which turns out to depend on the mass, angular momentum and the distance between the centers of the black holes. This last quantity turns out to be a genuine thermodynamic variable as a result of imposing the consistency of the fundamental equation in the context of GTD. Then, we use the thermodynamic curvature of the equilibrium space to determine the curvature singularities which, according to GTD, represent second order phase transitions. In addition, we show that the curvature singularities can also be interpreted as phase transitions from the point of view of classical thermodynamics.

This paper is organized as follows. In section \ref{sec:binsys}, we present the main aspects of the black hole binary system and find the conditions for the existence of horizons. In Sec. \ref{sec:gtd}, we present the metric of the phase space ${\cal T}$, which is invariant with respect to
total Legendre transformations and depends on the contact structure of ${\cal T}$. Then, in Sec. \ref{sec:equ}, we use the contact structure of ${\cal T}$ to study the behavior of the most relevant thermodynamic variables. Sec. \ref{sec:phases} is devoted to the study of the metric of the equilibrium space ${\cal E}$ and the associated curvature singularities which determine the phase transition structure of the binary system. Finally, in
Sec. \ref{sec:con}, we present the conclusions of our work.


\section{The binary system}
\label{sec:binsys}

The assumption that a binary system is composed of two stationary rotating objects reduces the complexity of the problem. In addition, one can assume that the system is located along an axis which coincides with the rotational axis of both compact objects. These assumptions allows us to investigate the corresponding gravitational field by using stationary axisymmetric spacetimes, which are described by the line element
\be
ds^2 = - f(dt-\omega d\varphi)^2 + f^{-1} \left[e^{2\gamma} (d\rho^2 d z^2) + \rho^2 d\varphi^2 \right] \ .
\ee
The corresponding vacuum field equations can be written in the Ernst representation as \cite{solutions}
\be
({\cal E} + \bar{\cal E})({\cal E} _{\rho\rho} + \rho^{-1} {\cal E} _\rho + {\cal E} _{zz}) = 2 ({\cal E} _\rho^2 + {\cal E} _z^2)\ ,
\label{ernsteq}
\ee
where the bar denotes complex conjugation, the subscripts represent partial derivatives, and  ${\cal E} = f+ i \Omega$ with
\bea
\omega_\rho &=& -\rho f^{-2} \Omega_z            \ , \quad \omega_z = \rho f^{-2} \Omega_\rho \ ,\\
4 \gamma_\rho &=&  \rho f^{-2}(f_\rho^2 - f_z^2 +\Omega_\rho^2 - \Omega_z^2 ) \ , \\
2 \gamma_z &=&  \rho f^{-2}(f_\rho f_z + \Omega_\rho \Omega_z) \ .
\eea
The Ernst equation (\ref{ernsteq}) can be solved by using Sibgatullin's method \cite{sib} which generates soliton-like solutions from the values of the Ernst potential on the axis of symmetry, i.e., ${\cal E}(\rho=0,z)=e(z)$. For the binary system composed of two identical Kerr black holes, the starting Ernst potential on the axis can be chosen as a particular case of the 2-soliton solution \cite{mr17}
\be
e(z) = \frac{z^2 -2 (M+ia)  z + c + i d }{z^2 - (M-ia) z + c-id }\ ,\quad e(z)+\bar e (z)=0 \ ,
\ee
where $M$ is the mass, $a=J/M$ is the specific angular momentum,
\bea
c & = & -\frac{1}{4}R^2 + 2M^2 - 2 a^2 - \sigma^2 \ , \\
d  & = & \sqrt{(R^2-4M^2+4a^2)(\sigma^2- M^2+a^2)}  \ .
\eea
The constant $R$ has been interpreted as the relative distance between the centers of the black holes and
\be
\sigma = \sqrt{M^2-\frac{J^2(R-2M)}{M^2(R+2M)}\ .
}
\ee

The Ernst potential for the entire spacetime can be obtained from the Sibgatullin integral
\be
{\cal E}(\rho, z) = \frac{1}{\pi}\int_{-1}^1 \frac{\mu(\zeta) e(\xi) d\zeta}{\sqrt{1-\zeta^2} }
\ee
where the unknown function $\mu(\zeta)$ satisfies the integral equation
\be
-\hskip-.4cm\int_{-1}^1 \frac{\mu(\zeta)[ e(\xi) \tilde e (\eta)] d\zeta }{(\xi-\eta) \sqrt{1-\zeta^2} } =0\ ,
\ee
with the normalization condition
\be
\frac{1}{\pi} \int_{-1}^1 \frac{\mu(\zeta)d\zeta}{\sqrt{1-\zeta^2}} = 1 \ .
\ee
Here, $\xi= z+ i\rho \zeta$, $\eta = z+i\rho \tau$ with $\tau\in [-1,1]$, $\tilde e (\eta) = \bar e (\bar\eta)$, and $-\hskip-.3cm\int$ represents the principal value integral. The computation of the Ernst potential ${\cal E}$ and the metric functions $f$, $\omega$
and $\gamma$ is laborious but straightforward. The explicit form of these functions can be written in several equivalent representations
\cite{Cabrera,Macias,mr17}. For the purpose of this work, however, the explicit form of the metric is not necessary.

An important property of this binary system is that along the axis
of symmetry and between the black holes there exists a
singularity, which has been  interpreted as corresponding to a
massless strut and is responsible for maintaining the black holes
apart. This is the reason why the binary system can be described
by means of a stationary spacetime. Another important
characteristic of this binary system is that the parameters of
each black hole satisfy the Smarr formula \cite{Smarr} 
\be M=
\frac{\kappa}{4\pi} S + 2 \Omega_H J \ , \label{smarr} \ee where
$\kappa$ is the  surface gravity of the black hole \be \kappa
=\frac{R(R+2M)\sigma }{4 \Big[M(R+3M)(R+2M)(M+\sigma)+2J^2 \Big]}\
, \ee $S$ is the entropy \be \label{fundamental2} S=8\pi M \left(
M+\sqrt{M^2-\frac{J^2(R-2M)}{M^2(R+2M)}}
\right)\left(1+\frac{2M}{R}\right)\,, \ee 
and $\Omega_H$ is the
angular velocity on the horizon 
\be \Omega_H
=\frac{J(R-2M)(R+2M)}{2M\Big[M(R+3M)(R+2M)(M+\sigma)+2J^2\Big]}\ .
\ee

The Smarr formula has been interpreted as generating the first law of black hole thermodynamics \cite{bch73,hawking76} in Einstein-Maxwell theory, in which black hole solutions are regular outside the horizon.  In the case of a binary system, the presence of a singularity along the axis of symmetry seems to be an obstacle for the validity of the Smarr formula and, consequently, of the first law of thermodynamics.
It is, therefore, remarkable that the Smarr formula holds for each of the constituents of the binary system. Nevertheless, a more detailed
analysis is necessary, involving the investigation of the corresponding Euclidean action and its analytic continuation \cite{hawking76}, in order to prove that in the case of a binary system the Smarr formula leads to a genuine first law of thermodynamics \cite{hawking76}.  In this work, we follow an alternative approach, in which we assume the validity of the Smarr formula in the thermodynamic context and analyze the properties of the binary system from the point of view of geometrothermodynamics. If the resulting thermodynamic system is physically reasonable, we interpret this as an indication of the thermodynamic validity of the Smarr formula.

\begin{figure}[ht]
\includegraphics[scale=0.3]{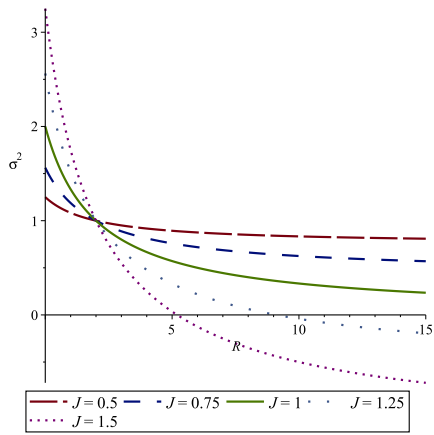}
\caption{Plot of $\sigma^2$ as a function of the distance $R$ for $M=1$ and several values of $J$.
Only the values included in the positive sector of the graphic are allowed for black hole binary systems. }
\label{figsigma2}
\end{figure}

Consequently, we assume that the entropy relation (\ref{fundamental2}) represents the fundamental thermodynamic equation for the binary system.
Notice that the entropy is a well-defined real function only if the condition
\be
\sigma^2 = M^2-\frac{J^2(R-2M)}{M^2(R+2M)} \geq 0
\ee
is satisfied, which is also the condition for the existence of the black hole horizon. For a fixed value of $M$ and $J$, there is maximum distance $R$ for which the black system exists. This behavior is illustrated in Fig. \ref{figsigma2}.

Since the entropy does not depend on the sign of $J$, let us consider $J>0$ for concreteness. Then, for $J\leq M^2$, the function $\sigma^2$ is always positive definite and the black hole binary system exists for any values of $R$.
In the limiting case $J=M^2$, the entropy becomes
\be
S(M,R)= 8\pi \frac{M^2}{R}\left[\left(\sqrt{R+2M} + \sqrt{M}\right)^2 - M^2\right] \ .
\ee
For values $J>M^2$, the function $\sigma^2$ is positive definite only in the interval (cf. Fig. \ref{figsigma2})
\be
0<R\leq R_{max} = \frac{2M(M^4 + J^2)}{J^2-M^4}  \ ,
\ee
and the entropy at the maximum distance reaches the value
\be
S= 16 \pi \frac{M^2J^2}{M^4+J^2}\ .
\ee
For larger values of the distance $R$, the horizon condition is not satisfied and the black hole binary system cannot exist. Thus, we see that the properties of the binary system depend not only on the relation between the mass and angular momentum of the rotating objects, but also on their relative distance. In the  limiting case $R\rightarrow \infty$, which physically corresponds to moving one of the black holes to spatial infinity, the entropy (\ref{fundamental2}) coincides with the fundamental equation of the Kerr black hole, if $J\leq M^2$.


\section{Geometrothermodynamic properties}
\label{sec:gtd}

In GTD, as in classical thermodynamics, to describe a thermodynamic system, it is necessary to know explicitly the corresponding fundamental equation and its behavior with respect to rescalings of the independent variables. Therefore, we first should establish the thermodynamic character of the parameters entering the fundamental equation. Indeed, it has been argued in \cite{qqs17,qqs18} that to correctly describe the thermodynamic properties of black holes, it is necessary to impose the condition that the corresponding fundamental equation be a quasi-homogeneous function. In the case of the function $S=S(M,J)$ given in
Eq.(\ref{fundamental2}), this means that the rescaling $M\rightarrow \lambda^{\beta_M} M$ and $J\rightarrow \lambda^{\beta_J} J$, where $\lambda$ and the $\beta$'s are real constants, should satisfy
the condition $S(\lambda^{\beta_M} M,\lambda^{\beta_J} J) = \lambda^{\beta_S} S(M,J)$. It is easy to see that this condition cannot be satisfied for any values of $\lambda$, $\beta_M$, $\beta_J$, and $\beta_S$. However, if we consider $R$ also as a thermodynamic variable, which rescales as $R \rightarrow \lambda^{\beta_R}R$, the quasi-homogeneity condition
$S(\lambda^{\beta_M} M,\lambda^{\beta_J} J, \lambda^{\beta_R}R) = \lambda^{\beta_S} S(M,J,R)$ holds if the relationships
\be
\beta_M = \beta_R= \frac{1}{2} \beta_J =\frac{1}{2}\beta_S\ .
\ee
are fulfilled.

The above result is important to apply the formalism of GTD in this particular case. Indeed, since the binary system is described by three extensive variables $M$, $J$ and $R$  plus the thermodynamic potential $S$, the corresponding phase space ${\cal T}$
is 7-dimensional with coordinates
$\{S,M,J,R, I_1,I_2,I_3\}$, where $I_a, \ a=1,2,3$ are  the intensive variables dual to the extensive variables $E^a= \{M,J,R\}$, respectively. The phase space is a contact Riemannian manifold $({\cal T}, \Theta, G^{^{II}})$, where $\Theta = dS - I_a dE^a$ is the canonical
contact 1-form and  the metric $G^{^{II}}$ is invariant with respect to total Legendre transformations and has been especially chosen to agree with the quasi-homogeneity property of black holes \cite{qqs17}. Then, we have
\bea
G^{^{II}} & = & \Theta^2 + (\beta_M M I_1 + \beta_J J I_2+ \beta_R R I_3)(- dM dI_1 + dJ dI_2 + dR dI_3) \ , \\
\Theta & = & dS - I_1 dM - I_2 d J - I_3 d R \ .
\label{gup}
\eea

The equilibrium space ${\cal E}$ with coordinates $\{E^a\}$ is defined by means of the smooth embedding map
$\varphi: {\cal E} \rightarrow {\cal T}$, or in coordinates $\varphi: \{E^a\}\mapsto \{S(E^a), E^a, I_a(E^a)\}$,
  such that
\be
\varphi^*(\Theta)=0 \ .
\label{flaw}
\ee
This condition guarantees that the equilibrium space ${\cal E}$ is mathematically well defined as a subspace of the phase space
${\cal T}$. Moreover, the assumption that the smooth map $\varphi$ exists, implies that the fundamental equation $S=S(E^a)$
must be known explicitly.

\subsection{Equilibrium conditions}
\label{sec:equ}

The vanishing of the contact 1-form $\Theta$ on ${\cal E}$ is an essential condition for the consistent formulation of GTD. Moreover, it induces the first law of thermodynamics and the equilibrium conditions, which must be satisfied on ${\cal E}$ and determine explicitly the behavior of all the thermodynamic variables. Indeed, the condition (\ref{flaw})
 is equivalent to the first law of black hole thermodynamics, i.e.,
\be
dS = \frac{1}{T} dM - \frac{\Omega_H}{T} d J - \frac{\cal F}{T} d R  \ .
\ee
In this way, we identify the intensive variables $I_a$ which are uniquely defined by the equilibrium conditions
\be
\frac{1}{T} = \frac{\partial S}{\partial M} \ , \quad
\frac{\Omega_H}{T} = - \frac{\partial S}{\partial J} \ , \quad
\frac{\cal F}{T}  = - \frac{\partial S}{\partial R} \ ,
\ee
in terms of the fundamental equation $S=S(M,J,R)$. Here, we have introduced the ``force'' ${\cal F}$ as the dual to the distance $R$. These expressions allow us to compute the explicit form of the corresponding intensive variables:
\bea
\label{Temperature} T & = & \frac{R(R+2M)\sigma
}{16\pi\Big[M(R+3M)(R+2M)(M+\sigma)+2J^2 \Big]}\,, \\
\Omega_H &=&\frac{J(R-2M)(R+2M)}{2M\Big[M(R+3M)(R+2M)(M+\sigma)+2J^2\Big]}\,, \\
\mathcal{F}&=&
\frac{M\Big[M^2(R+2M)(M+\sigma)+2J^2\Big]}{R\Big[M(R+3M)(R+2M)(M+\sigma)+2J^2\Big]}\,.
\eea
As before, all these quantities are well defined only if $\sigma>0$. In the limit $R\rightarrow \infty$, they all reduce to the case of a single Kerr black hole:
\bea
T^K & = & \frac{\sigma }{16\pi M(M+\sigma) }\,,  \quad  \sigma=\left(M^2-\frac{J^2}{M^2} \right)^{\frac{1}{2}}\,\\
\Omega^K_H &= &\frac{J }{2M( M^2+\sqrt{M^4-J^2})}\,, \\
\mathcal{F}^K & = & 0\, .
\eea

\begin{figure}[ht]
\includegraphics[scale=0.45]{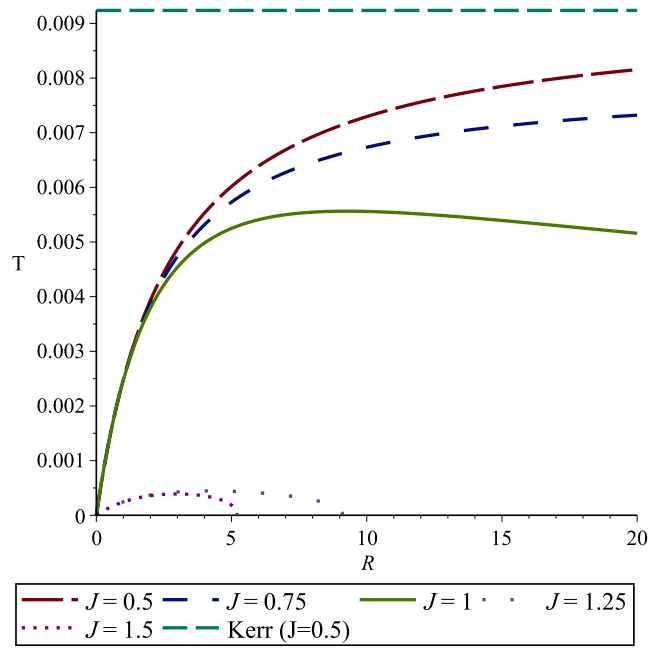}
\caption{
\label{fig:temp}
Temperature  $T$  as a function of  $R$  with fixed mass ($M=1$) and different values for the angular momentum.}
\end{figure}

The behavior of the temperature in terms of the distance $R$ is shown in Fig. \ref{fig:temp} for a fixed value of the mass and different values of the angular momentum.
We see that in general the temperature of the binary system is less than the temperature of a single Kerr black hole for a
fixed value of $J$. Consequently, in a binary system, the presence of a black hole affects the properties of its companion by reducing its temperature.

\begin{figure}[ht]
\includegraphics[scale=0.45]{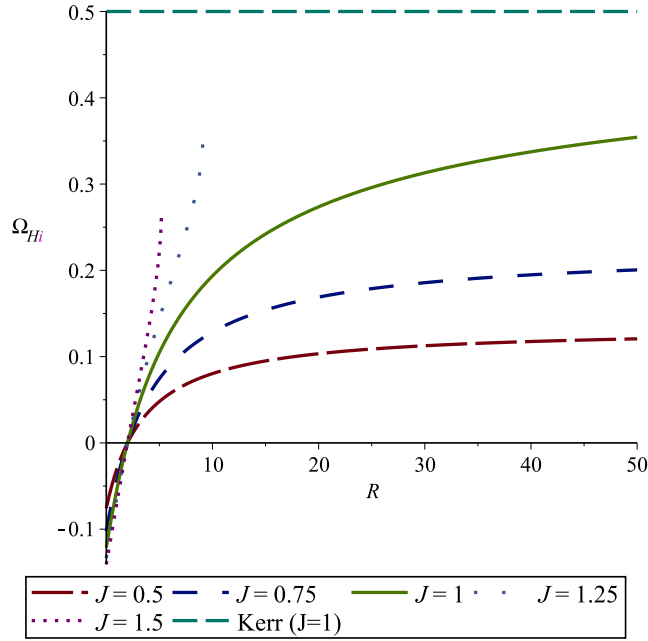}
\caption{
\label{fig:omega}
Angular velocity $\Omega_H$  as a function of  $R$  with $M=1$ and different values for the angular momentum.}
\end{figure}

As for the angular velocity, its behavior is illustrated in Fig. \ref{fig:omega}.
The maximum angular velocity on the horizon for a fixed value of $J$ corresponds to the Kerr limit ($R\rightarrow\infty$). As $R$ decreases,  the angular velocity decreases as well, reaching the value $\Omega_H=0$ for $R=2M$, independently of the value of the angular momentum. From this point backwards the angular velocity changes its sign. This remarkable effect is due to presence of the second black hole at a very short distance.

\begin{figure}[ht]
\includegraphics[scale=0.45]{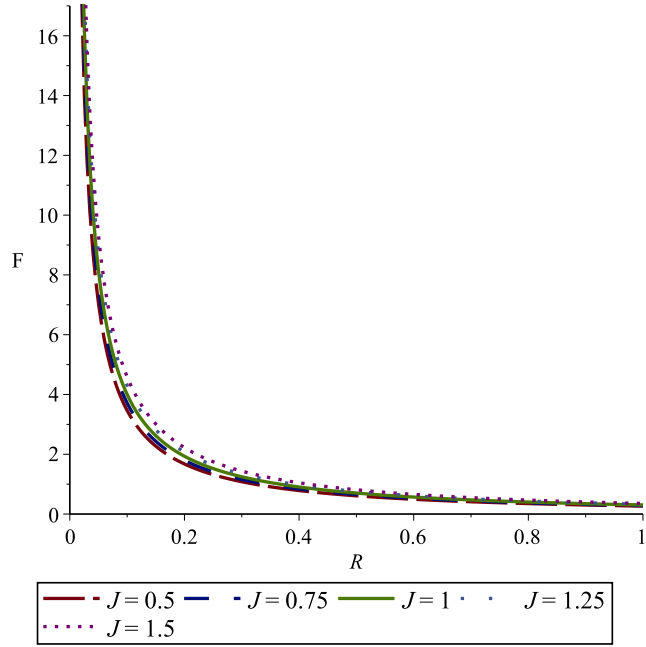}
\caption{
\label{fig:force}
The force  ${\cal F}$ as a a function of  $R$  with $M=1$ and different values for the angular momentum.}
\end{figure}

In Fig. \ref{fig:force}, we present the plot of the effective force ${\cal F}$ as a function of the distance $R$.
The behavior is very similar for all the values of the angular momentum and all the curves tend towards zero, which is the limiting value of the Kerr black hole. Moreover, as the distance between the black holes decreases, the force increases exponentially and diverges asymptotically for vanishing distances.

\subsection{Phase transitions}
\label{sec:phases}

Phase transitions are usually associated with severe changes inside the thermodynamic system, which imply  the  breakdown of the equilibrium conditions, i.e., during phase transitions classical equilibrium thermodynamics is not valid anymore. The formalism of GTD incorporates this fact in an analog way and assumes that a phase transition should correspond to a breakdown of the geometric structure of the equilibrium space ${\cal E}$, which occurs in particular when curvature singularities are present. Therefore, to investigate the phase transition structure of a thermodynamic system, in GTD we investigate the curvature singularities of the corresponding equilibrium space ${\cal E}$.
The geometric properties of ${\cal E}$ are dictated by the corresponding Riemannian metric $g$, which is induced canonically by the pullback
 $\varphi^*(G)=g$ of the metric $G$ of ${\cal T}$. For the particular case of a black hole binary system, the metric of the phase space is  given in Eq.(\ref{gup}) and, consequently, for the equilibrium space we obtain
\bea
\label{gtdBS}
g^{^{II}}=\beta_S {S}\Big(-\frac{\partial^2
S}{\partial M^2 }dM^2 +2\frac{\partial^2 S}{\partial J \partial R
}dJ dR+\frac{\partial^2 S}{\partial J^2 }dJ^2+\frac{\partial^2
S}{\partial R^2 }dR^2\Big)\,,
\eea
where $S(M,J,R)$ is the fundamental equation (\ref{fundamental2}) of the binary system, from which the explicit expressions of the metric components can be written as follows:
\bea
g^{^{II}}_{MM} = && -\frac{16\pi\beta_S S}{\sigma^3 M^3 R(R+2M)^3}
 [ M^3 (M^3+\sigma^3) (R+6M)(R+2M)^3 \nonumber \\
&& - M^2J^2 (R+2M)^2(20M^2 - 8MR-3R^2) - 2 R^2 J^4 ]\ ,\\
g^{^{II}}_{JR} = &&-\frac{16\pi \beta_S J S [M^3(R+2M)^2 - 2 J^2 (R-2M)]}{\sigma^3 M R^2(R+2M)^2} \ , \\
g^{^{II}}_{JJ} = && - \frac{8\pi \beta_S M S (R-2M)}{\sigma^3 R} \ , \\
g^{^{II}}_{RR} = && \frac{32\pi\beta_S S }{\sigma^3 M R^3 (R+2M)^3} [ M^6(M+\sigma)(R+2M)^3 +  J^4(8M^2-3R^2) \nonumber \\
&& + J^2 M^2 (R+2M)^2 [\sigma(2M-R) + 4M^2 - MR] \,] \ .
\eea

\begin{figure}[ht]
\includegraphics[scale=0.35]{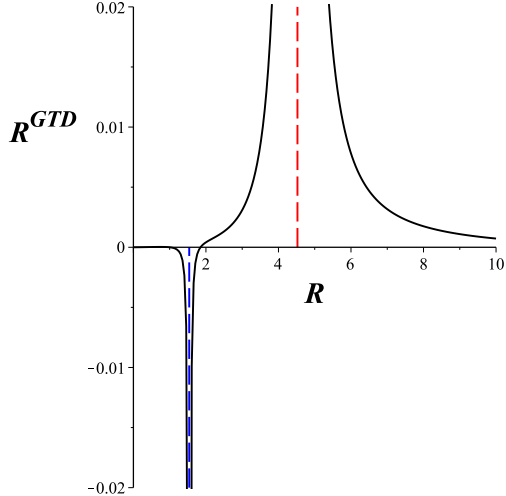}
\caption{Curvature
scalar $R^{GTD}$ as a function of the distance between the black holes. The mass and angular momentum have the fixed values
$M=1$ and $J=1.1$, respectively.
\label{fig:curv}}
\end{figure}

The corresponding curvature scalar can be represented as
\be
R^{^{GTD}}= \frac{N(M,J,R)}{ D_1 D_2^2 D_3^3}\ ,
\ee
where
\be
D_1= 256 \beta_S \pi^2 \sigma M^6 (M+\sigma)^3 (R+2M)^3 \ ,
\ee
\be
D_2 = M^3 (M^3+\sigma^3) (R+6M)(R+2M)^3  - M^2J^2 (R+2M)^2(20M^2 - 8MR-3R^2) - 2 R^2 J^4 \ ,
\label{d2}
\ee
\be
D_3 = (M+\sigma) M^4 (R-2M)(R+2M)^3 + J^2[M^2(3R-2M)(R+2M)^2 - 4J^2(R-2M)] \ ,
\label{d3}
\ee
and $N(M,J,R)$ represents the numerator of the curvature scalar. Clearly, the curvature singularities are determined by the zeros of $D_2$ and $D_3$.
For a fixed value of the mass black hole, the location of these singularities depend on the values of the angular momentum and the distance.
A particular example of this singular behavior is illustrated in Fig. \ref{fig:curv}.

\begin{figure}[ht]
\includegraphics[scale=0.35]{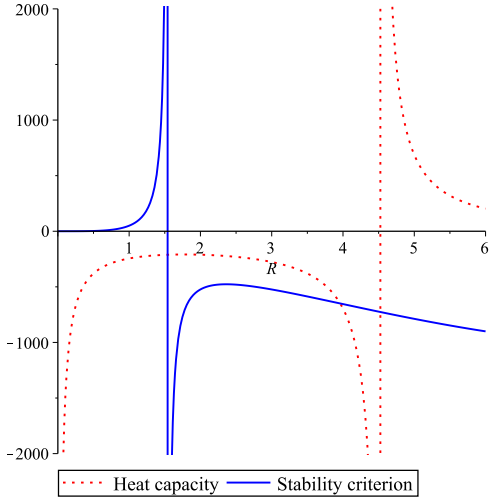}
\caption{The heat capacity $C$ and stability condition  $S_c^{-1}$ as  functions of $R$ for $M=1$, $J=1.1$, which are the same values as in Fig. \ref{fig:curv}. The divergence of $C$ ($S_c^{-1}$) coincides with the second (first)  singularity of $R^{^{GTD}}$ located at $R\approx 4.52$
($R\approx 1.54$). The scale of  $S_c^{-1}$ has been modified to  visualize the effect near the singularity.
\label{fig:heat}}
\end{figure}

According to GTD, the black hole binary system undergoes second order phase transitions at those points of the equilibrium space where any of the conditions $D_2=0$ or $D_3=0$ are met. This result is invariant in the sense that it does not depend on the choice of coordinates because it is based on the analysis of a scalar quantity.
On the other hand, in classical thermodynamics the phase transition structure of a system is determined by the behavior of the response functions, according to the Ehrenfest scheme, and the stability criteria \cite{callen}. However, it has been determined that sometimes the predictions of the  Ehrenfest scheme do not coincide with the observed phase transitions \cite{jae98}. We believe that the formalism of GTD with its invariant definition of phase transitions could shed some light on this issue. Nevertheless, in the case of the binary system under consideration in this work, it is possible to find a relation between the GTD and the Ehrenfest formalisms. Indeed, consider the heat capacity
\be
C= T\frac{\partial S}{\partial T} = - \frac{\left(\frac{\partial S}{\partial M}\right)^2}{\frac{\partial^2 S}{\partial M^2}} \ ,
\ee
which is evaluated at constant $J$ and constant $R$. It is then straightforward to compute this response function and it turns out that that $C\sim 1/D_2$, where $D_2$ is the polynomial defined in Eq.(\ref{d2}). This means that the zeros of $D_2$, which are associated with curvature singularities in the equilibrium space, correspond to divergencies of the heat capacity as well. To illustrate the behavior of the heat capacity, we plot a particular case in Fig. \ref{fig:heat}.

A stability condition which is usually applied in classical thermodynamics relates the derivatives of the fundamental equation and can expressed as \cite{callen}
\be
S_c = \left( \frac{\partial^2 S}{\partial J^2}\right) \left( \frac{\partial ^2 S}{\partial R^2}\right)
 - \left(\frac{\partial^2 S}{\partial J \partial R}\right)^2 >0\ .
\ee
If  $S_c$ happens to be negative, the system is unstable. The points where $S_c=0$ are interpreted as corresponding to changes in the stability properties of the system and are usually accompanied by phase transitions. A straightforward computation of this stability condition for the binary system (\ref{fundamental2}) shows that $S_c\sim  D_3$. In Fig. \ref{fig:heat}, we illustrate the behavior of
$S_c^{-1}$, showing a divergence exactly at the point where the first curvature singularity is located.

We thus conclude that the curvature singularities detected with the GTD formalism in the equilibrium space of a black hole binary system correspond to second order phase transitions, which are accompanied by modifications in the stability properties of the system.

\section{Final remarks}
\label{sec:con}

In this work,  we studied a binary system composed of two identical co-rotating Kerr black holes. It is assumed that the system is stationary and possesses a spatial   symmetry with respect to an axis, which coincides with the rotational axis of the sources.
For a fixed value of the angular momentum, it turns out that the distance between the centers of the sources determines whether the system is composed of two black holes or two naked singularities. We limit ourselves to the study of black holes. In this case, the Smarr formula represents a relationship between the physical parameters, which determine the gravitational field of the binary system, and we assume that it can be interpreted as determining the fundamental thermodynamic equation.

We investigate the thermodynamic properties of the system by using entirely the formalism of GTD. First, we impose the condition that the  fundamental equation corresponds to a quasi-homogeneous function of degree different from one. As a consequence, we show that
the distance $R$ must be considered as a thermodynamic variable. This is an important result because the presence of $R$ at the thermodynamic level drastically changes the behavior of the binary system. This also increases the dimensionality of the phase space
${\cal T}$, which is a contact manifold endowed with a Legendre invariant Riemannian metric. This kind of extended thermodynamics has been investigated very actively in the past years, by considering the cosmological constant as a  thermodynamic variable \cite{kmm17}. In the case analyzed in this work, the distance $R$ plays the role of additional thermodynamic variable, which is responsible for increasing the dimensionality of the thermodynamic phase space, leading to another version of extended thermodynamics.

We use the contact structure of  ${\cal T}$ and
its projection on the equilibrium space ${\cal E}$, to impose the first law and the equilibrium conditions from which all the relevant thermodynamic variables can be obtained.  In particular, we study the behavior of the temperature, the angular velocity on the horizon, and the force which is dual to the distance $R$. The common feature is that the presence of an accompanying black hole reduces the temperature and the angular velocity of the partner and that asymptotically ($R\rightarrow \infty$) the system reduces to that of a single Kerr black hole.

To study the phase transition structure of the binary system, we use the fact that, according to GTD, the curvature singularities of the equilibrium space correspond to second order phase transitions. We compute the curvature scalar of the equilibrium space and show that in general there are two singularities. The existence of the corresponding phase transitions is corroborated by using the Ehrenfest scheme and the stability condition of classical thermodynamics. In fact, we show that the first singularity can be interpreted as due to the breakdown of the stability condition and the second one corresponds to a divergence at the level of the  heat capacity.

Our results show that the binary system of two identical Kerr black holes can be interpreted as a thermodynamic system in which the distance between the black holes must be considered as a thermodynamic variable that strongly affect the physical properties of the system. We interpret this result also as an indication of the thermodynamic validity of the Smarr formula in the framework of a binary system.

\section*{Acknowledgments}

This work was carried out within the scope of the project CIAS 2312
supported by the Vicerrector\'\i a de Investigaciones de la Universidad
Militar Nueva Granada - Vigencia 2017.
 This work was partially supported  by UNAM-DGAPA-PAPIIT, Grant No. 111617, and by the Ministry of Education and Science of RK, Grant No.
BR05236322 and AP05133630.



\begin{thebibliography}{99}

\bibitem{Cabrera} I. Cabrera--Mungu\'\i a, V. E. Ceron, L. A. L\'opez, and  O. Pedraza,
Phys. Lett.  B  {\bf 772}, 10 (2017).


\bibitem{Majumdar} S. D. Majumdar, Phys. Rev. {\bf 72}, 390 (1947).

\bibitem{Papapetrou} A. Papapetrou,  Proc. Roy. Irish Acad. A {\bf 51}, 191 (1947).


\bibitem{Israel1} W. Israel and G. A. Wilson, J. Math. Phys. {\bf 13}, 865 (1972).

\bibitem{Perjes} Z. A. Perjes, Phys. Rev. Lett. {\bf 27}, 1668 (1971).

\bibitem{Parker} L. Parker, R. Ruffini,  and D. Wilkins,
Phys. Rev. D {\bf 7}, 2874 (1973).

\bibitem{Varzugin} G. G. Varzugin, Theor. Math. Phys. {\bf 116}, 1024 (1998).


\bibitem{Komar} A. Komar, Phys. Rev. {\bf 113}, 934 (1959).

\bibitem{Macias} I. Cabrera--Mungu\'\i a, C. L\"ammerzahl , and A. Mac\'\i as, arXiv:1309.2556v2 [gr--qc]


\bibitem{Miguel} M. S. Costa, C. Herdeiro, and C. Robelo,
Phys. Rev. D {\bf 79}, 123508 (2009).



\bibitem{Smarr} L. Smarr, Phys. Rev. Lett. {\bf 30}, 71 (1973).


\bibitem{Bekenstein} J. D. Bekenstein,
Phys. Rev. D {\bf 7}, 2333 (1973).



\bibitem{bch73} J. M. Bardeen, B. Carter and S. W. Hawking, Commun. Math. Phys. {\bf 31}, 161 (1973).

\bibitem{hawking76}  S. W. Hawking, Phys. Rev. D {\bf 13}, 191 (1976).



\bibitem{weibook} F. Weinhold, {\it Classical and Geometrical Theory of Chemical and Phase
Thermodynamics} (Wiley, Hoboken, New Jersey, 2009).



\bibitem{ruprev} G. Ruppeiner, Springer Proc. Phys. {\bf 153}, 179 (2014).



\bibitem{quev07} H. Quevedo, 
                 J. Math. Phys. {\bf 48}, 013506 (2007).


\bibitem{solutions} H. Stephani, D. Kramer, M. MacCallum, C. Hoenselaers,
and E. Herlt, {\em Exact solutions of Einstein's field equations},
(Cambridge University Press, Cambridge, UK, 2003)



\bibitem{sib} N. R. Sibgatullin,
{\it Oscillations and waves in strong gravitational and electromagnetic waves} (Springer-Verlag, New York, 1991)



\bibitem{mr17} V.S. Manko and E. Ruiz, Phys. Rev. D {\bf 96}, 104016  (2017).


\bibitem{qqs17} H. Quevedo, M. N. Quevedo, and A. S\'anchez,
Eur. Phys. J. C {\bf 77}, 158 (2017).


\bibitem{qqs18} H. Quevedo, M. N. Quevedo, and A. S\'anchez,
Eur. Phys. J. C {\bf 79}, 229 (2019).

\bibitem{callen} H. B. Callen, {\em Thermodynamics and an introduction to
thermostatics} (John Wiley  Sons, Inc., New York, 1985).



\bibitem{jae98}
G. Jaegger,
{\it The Ehrenfest Classification of Phase Transitions: Introduction and Evolution},
Archive for History of Exact Sciences {\bf 53}, 51 (1998).

\bibitem{kmm17} D. Kubiznak, R. B. Mann, and  M.  Teo,
Class. Quant. Grav. {\bf 34}, 063001  (2017).

\end{thebibliography}
\end{document}